\documentclass[aps,twocolumn,floatfix,showpacs,amsmath,amsfonts]{revtex4}

\usepackage{graphicx}
\usepackage{bm}
\usepackage{units}
\usepackage{color}
\usepackage{upgreek}
\usepackage{gensymb}

\begin{document}

\title{Exciton-phonon interaction breaking all antiunitary symmetries in external magnetic fields}

\author{Frank Schweiner}
\author{Patric Rommel}
\author{J\"org Main}
\author{G\"unter Wunner}
\affiliation{Institut f\"ur Theoretische Physik 1, Universit\"at Stuttgart,
  70550 Stuttgart, Germany}
\date{\today}

\begin{abstract}
Recent experimental investigations by M.~A{\ss}mann 
\emph{et al.}~[Nature Mater.~\textbf{15}, 741 (2016)] on the spectrum of magnetoexcitons
in cuprous oxide revealed the statistics of a Gaussian unitary ensemble
(GUE). 
The model of F.~Schweiner \emph{et al.}~[Phys.~Rev.~Lett.~\textbf{118}, 046401 (2017)],
which includes the complete cubic valence band structure of the solid,
can explain the appearance of GUE statistics if the magnetic field is
not oriented in one of the symmetry planes of the cubic lattice. However, it
cannot explain the experimental observation of GUE statistics 
for all orientations of the field. 
In this paper we investigate the effect of 
quasi-particle interactions or especially the exciton-phonon interaction on the
level statistics of magnetoexcitons
and show that the motional Stark field induced by the exciton-phonon interaction leads
to the occurrence of GUE statistics for arbitrary orientations of the magnetic
field in agreement with experimental observations.
Importantly, the breaking of all antiunitary symmetries 
can be explained only by considering both the exciton-phonon interaction and
the cubic crystal lattice.
\end{abstract}


\maketitle

\section{Introduction}

Excitons are fundamental quasi-particles in semiconductors,
which are the elementary excitations of the electronic system.
Consisting of a negatively charged electron in the conduction
band and a positively charge hole in the valence band, which interact
via a screened Coulomb interaction, excitons are often
regarded as the hydrogen analog of the solid state.
Especially excitons in cuprous oxide $\left(\mathrm{Cu_{2}O}\right)$
are of interest due to their high Rydberg energy.
Only three years ago 
an almost perfect hydrogen-like absorption series
has been observed in $\mathrm{Cu_{2}O}$
up to a principal quantum number of $n=25$
by T.~Kazimierczuk~\emph{et~al.}~\cite{GRE}. 
This experiment has opened the field of research of giant Rydberg excitons
and has stimulated a large number of experimental and theoretical 
investigations~\cite{75,76,50,28,80,100,125,78,79,150,74,77,200,250},
in particular as concerns the level statistics 
and symmetry-breaking effects~\cite{QC,QC2,175,225}.

If symmetries are broken, the classical dynamics of a system
often becomes nonintegrable and chaotic. However, 
since the description of chaos by trajectories and
Lyapunov exponents is not possible in quantum mechanics,
classical chaos manifests itself in
quantum mechanics in a different way~\cite{QSC,QCI}. 
The Bohigas-Giannoni-Schmit conjecture~\cite{QC_1} suggests
that quantum systems
with few degrees of freedom and with a chaotic classical
limit can be described by random matrix theory~\cite{QSC_29,QSC_30}
and show typical level spacings.
If the classical dynamics is regular, the level spacing obeys Poissonian
statistics. At the transition to chaos,
the level spacing statistics 
changes to the statistics of
a Gaussian orthogonal ensemble (GOE), a
Gaussian unitary ensemble (GUE) 
or a Gaussian symplectic ensemble (GSE)
as symmetry reduction leads to a correlation of levels
and hence to a strong suppression of crossings~\cite{QSC}.
To which of the three universality classes, i.e., to the orthogonal, the unitary or
the symplectic universality class, a given system
belongs is determined by the remaining symmetries in the system.
While GOE statistics appears if there is at least one remaining
antiunitary symmetry in the system, for GUE statistics \emph{all}
antiunitary symmetries have to be broken. GSE statistics can be observed
for systems with \emph{time-reversal invariance 
possessing Kramer's degeneracy
but no geometric symmetry at all}~\cite{QSC}.

The hydrogen-like model of excitons is
often too simple to account for the large number of effects
due to the surrounding solid.
Some essential corrections to this model
comprise, e.g., the inclusion of the complete 
cubic valence band structure~\cite{28,100,17_17_18,17_17_26,7_11,17_17,17_15,44_12},
which leads to a complicated fine-structure splitting,
or the interaction with quasi-particles like
phonons~\cite{75,AP3,M1_7,2}.

An important experimental observation by 
M.~A{\ss}mann~\emph{et al.}~\cite{QC,QC2}, which cannot be
explained by the hydrogen-like model, is the 
appearance of GUE statistics 
for excitons in an external magnetic field in $\mathrm{Cu_{2}O}$.
This observation implies that all antiunitary symmetries
are broken in the system. However, for most of the physical systems still there is 
at least one antiunitary symmetry left~\cite{QC_5,QC_16,QSC_12,QSC_13,QSC_15,QSC_16,QSC_17,QSC_18,QC_3,QC_4}.
This also holds for atoms in constant external fields~\cite{QSC_19,QC_2,GUE1}.
Hence, based on the hydrogen-like model one would expect to 
observe the statistics of a Gaussian 
orthogonal ensemble (GOE).

As an explanation, M.~A{\ss}mann~\emph{et al.}~\cite{QC,QC2}
attributed the breaking of all antiunitary symmetries 
observed for magnetoexcitons to the 
interaction of excitons with phonons.
In a recent letter we have shown theoretically that the
combined presence of an external magnetic field and
the cubic valence band structure of $\mathrm{Cu_{2}O}$
is sufficient to break all antiunitary symmetries in the system
without the need for phonons~\cite{175}.
However, this breaking appears only if the magnetic
field is not oriented in one of the symmetry planes of the
cubic lattice of $\mathrm{Cu_{2}O}$.
Hence, our model cannot explain the fact that
GUE statistics has been observed in the experiment for all 
directions of the magnetic field~\cite{QC,QC2}.
This raises again the question about the influence of the
exciton-phonon interaction on the level spacing statistics of the
exciton spectra.

In this paper we will discuss in detail the effects which leads
to the appearance of GUE statistics whether or not the external fields
are oriented in one of the symmetry planes of the cubic lattice.
For fields oriented in a symmetry plane of the laatice, we explain
that the interaction of the exciton with other quasi-particles like phonons
is not able to restore the broken antiunitary symmetries. 
As regards the other orientations of the external fields,
we discuss that the exciton-phonon interaction 
leads to a finite momentum
of the exciton center of mass and thus to 
the appearance of a magneto Stark effect in
an external magnetic. The electric field connected to this effect
then causes in combination with the cubic lattice
the breaking of all antiunitary symmetries.
Hence, we explain the appearance of GUE statistics for all orientations
of the external fields.

The paper is organized as follows:
In Sec.~\ref{sub:Hamiltonian}
we discuss the Hamiltonian of excitons in $\mathrm{Cu_{2}O}$ 
when considering the complete valence band structure and the
presence of external fields. We explain how to solve the
corresponding Schr\"odinger equation numerically by using a complete
basis in Sec.~\ref{sub:Complete basis}. The calculation of the 
level spacing distributions is 
shortly presented in Sec.~\ref{sub:Level-spacing-distributions}.
We then show the breaking of all antiunitary symmetries 
in external fields. At first, we treat the case 
with the plane spanned by the external fields not being
identical to one symmetry plane of the lattice in Sec.~\ref{sec:analytical}. 
In Sec.~\ref{sec:phonons} we discuss the effect of the exciton-phonon
interaction and, hence, the motional Stark field, 
on the spectra if the external fields \emph{are} oriented in 
one of the symmetry planes of the lattice. 
We finally give a short summary and outlook in Sec.~\ref{sec:Summary}.

\section{Theory\label{sec:Theory}}

In this section we briefly introduce the Hamiltonian of excitons in
$\mathrm{Cu_{2}O}$ and show how to
solve the corresponding Schr\"odinger equation
in a complete basis. Furthermore, we discuss how to
determine the level spacing statistics of the exciton spectra numerically
and to which level spacing distribution functions the results will be compared.
For more details see Refs.~\cite{100,125,225,GUE4,75}
and further references therein.

\subsection{Hamiltonian\label{sub:Hamiltonian}}

When neglecting external fields, the Hamiltonian
of excitons in direct semiconductors is given by~\cite{17_17}
\begin{equation}
H=E_{\mathrm{g}}+V\left(\boldsymbol{r}_{e}-\boldsymbol{r}_{h}\right)+H_{\mathrm{e}}\left(\boldsymbol{p}_{\mathrm{e}}\right)+H_{\mathrm{h}}\left(\boldsymbol{p}_{\mathrm{\mathrm{h}}}\right)\label{eq:Hpeph}
\end{equation}
with the energy $E_{\mathrm{g}}$ of the band gap
between the lowest conduction band and the highest 
valence band. 
The Coulomb interaction between the electron (e) and the hole (h)
is screened by the dielectric constant~$\varepsilon$:
\begin{equation}
V\left(\boldsymbol{r}_{e}-\boldsymbol{r}_{h}\right)=-\frac{e^{2}}{4\pi\varepsilon_{0}\varepsilon}\frac{1}{\left|\boldsymbol{r}_{e}-\boldsymbol{r}_{h}\right|}.
\end{equation}

\begin{figure}
\begin{centering}
\includegraphics[width=0.8\columnwidth]{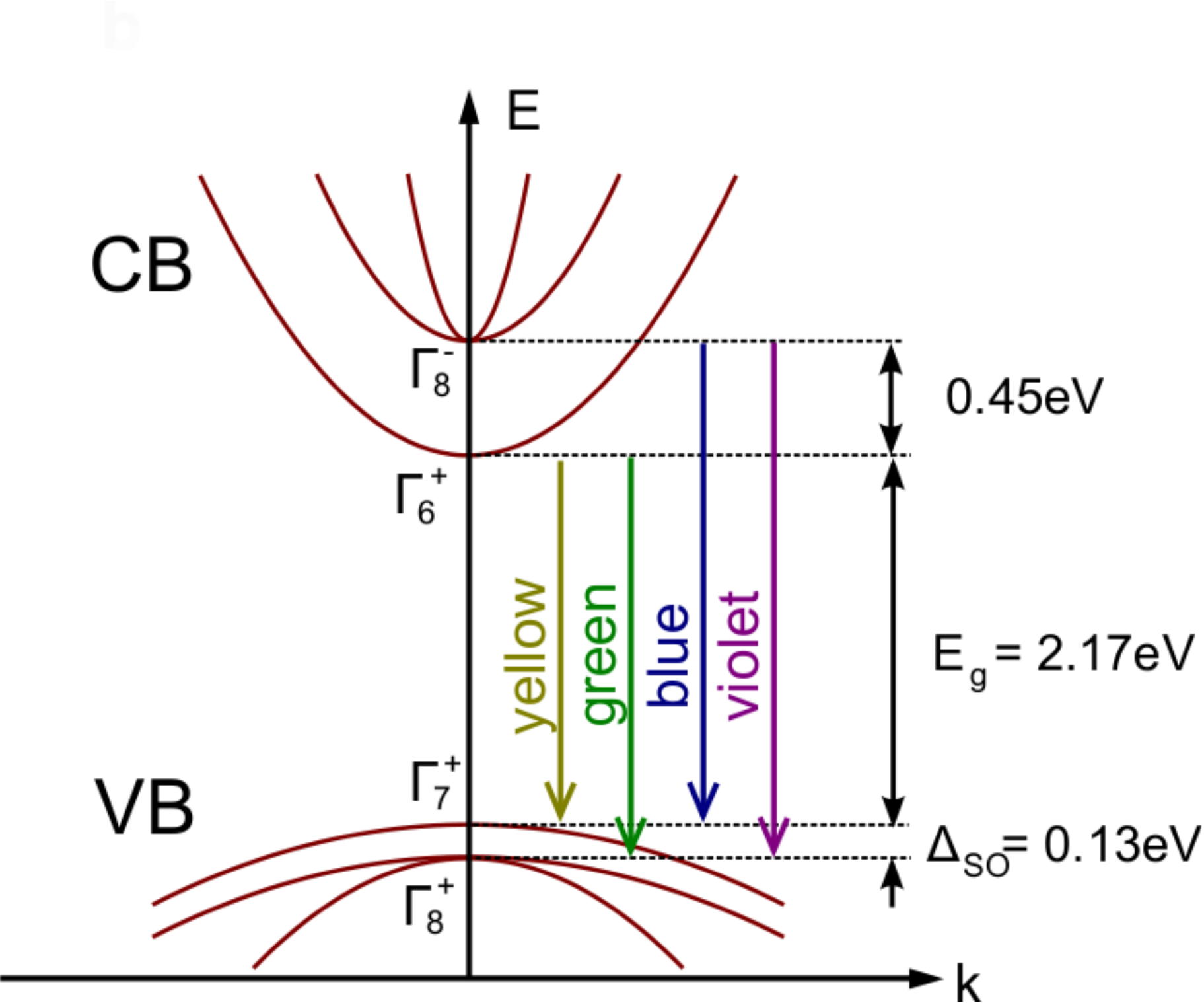}
\par\end{centering}

\protect\caption{Band structure of $\mathrm{Cu_{2}O}$~\cite{GRE}. 
As a consequence of the spin-orbit coupling
(\ref{eq:soc}) the valence band splits into a lower lying fourfold-degenerate 
band (including the hole spin) of symmetry $\Gamma_{8}^{+}$ of and a higher lying twofold-degenerate
band of symmetry $\Gamma_{7}^{+}$. 
The lowest lying conduction band of $\mathrm{Cu_{2}O}$
has $\Gamma_{6}^{+}$ symmetry.
Depending on the bands involved, one
distinguishes between the yellow, green, blue, and violet exciton series.
Due to the cubic symmetry of $\mathrm{Cu_{2}O}$, the symmetry of the
bands can be assigned by the irreducible representations $\Gamma_{i}^{\pm}$
of the cubic group $O_{\mathrm{h}}$, where the superscript $\pm$
denotes the parity. 
\label{fig:Band-structure-of}}

\end{figure}

\textcolor{blue}{Since the conduction band is
close to parabolic at zone center},
the kinetic energy of the electron is 
given by the simple expression
\begin{equation}
H_{\mathrm{e}}\left(\boldsymbol{p}_{\mathrm{e}}\right)=\frac{\boldsymbol{p}_{\mathrm{e}}^{2}}{2m_{\mathrm{e}}},
\end{equation}
with the effective mass $m_{\mathrm{e}}$ of the electron.
As regards the valence bands, the situation is more complicated.
In all crystals with zinc-blende and diamond structure 
the valence band is threefold degenerate at the center of 
the first Brillouin zone or the $\Gamma$ point~\cite{17_17,17_17_11}.
Due to the spin-orbit coupling~\cite{SST,SO},
the degeneracy is lifted in $\mathrm{Cu_{2}O}$ and two of the three valence bands are shifted
towards lower energies~\cite{20}.
This is shown in Fig.~\ref{fig:Band-structure-of}.
The competition between the dispersion of the threefold degenerate orbital valence band
with the spin-orbit splitting is responsible for a strong non-parabolicity of the valence bands.

The kinetic energy of a hole within these valence bands is given
by~\cite{44_12,80,100}
\begin{eqnarray}
H_{\mathrm{h}}\left(\boldsymbol{p}_{\mathrm{h}}\right) & = & H_{\mathrm{so}}+
\left(1/2\hbar^{2}m_{0}\right)\left\{ \hbar^{2}\left(\gamma_{1}+4\gamma_{2}\right)\boldsymbol{p}_{\mathrm{h}}^{2}\right.\phantom{\frac{1}{1}}\nonumber \\
 & + & 2\left(\eta_{1}+2\eta_{2}\right)\boldsymbol{p}_{\mathrm{h}}^{2}\left(\boldsymbol{I}\cdot\boldsymbol{S}_{\mathrm{h}}\right)\phantom{\frac{1}{1}}\nonumber \\
 & - & 6\gamma_{2}\left(p_{\mathrm{h}1}^{2}\boldsymbol{I}_{1}^{2}+\mathrm{c.p.}\right)-12\eta_{2}\left(p_{\mathrm{h}1}^{2}\boldsymbol{I}_{1}\boldsymbol{S}_{\mathrm{h}1}+\mathrm{c.p.}\right)\phantom{\frac{1}{1}}\nonumber \\
 & - & 12\gamma_{3}\left(\left\{ p_{\mathrm{h}1},p_{\mathrm{h}2}\right\} \left\{ \boldsymbol{I}_{1},\boldsymbol{I}_{2}\right\} +\mathrm{c.p.}\right)\phantom{\frac{1}{1}}\nonumber \\
 & - & \left. 12\eta_{3}\left(\left\{ p_{\mathrm{h}1},p_{\mathrm{h}2}\right\} \left(\boldsymbol{I}_{1}\boldsymbol{S}_{\mathrm{h}2}+\boldsymbol{I}_{2}\boldsymbol{S}_{\mathrm{h}1}\right)+\mathrm{c.p.}\right)\right\} \phantom{\frac{1}{1}}\label{eq:Hh}
\end{eqnarray}
with $\boldsymbol{p}=\left(p_1,\,p_2,\,p_3\right)$, $\left\{ a,b\right\} =\frac{1}{2}\left(ab+ba\right)$ and
c.p.~denoting cyclic permutation.
The three Luttinger parameters $\gamma_{i}$
as well as the parameters $\eta_{i}$ 
describe the behavior and the 
anisotropic effective mass of the hole in the vicinity of the $\Gamma$ point.
Note that the parameters $\eta_{i}$ are often much smaller than the Luttinger parameters
and are neglected in the following~\cite{17_17_18,80,100}.
\textcolor{black}{We have recently shown that the inclusion of quartic and higher-order terms in
$p$ in the kinetic energies of the electron and the hole is not necessary
due to their negligible size~\cite{200}.}

The quasispin $I=1$ describes the threefold degenerate valence band
and is a convenient abstraction to denote the three 
orbital Bloch functions $xy$, $yz$, and $zx$~\cite{25}.
The matrices $\boldsymbol{I}_j$
and $\boldsymbol{S}_{\mathrm{h}j}$
denote the three
spin matrices of the quasispin $I$
and the hole spin $S_{\mathrm{h}}=1/2$
while $\boldsymbol{I}$ and $\boldsymbol{S}_{\mathrm{h}}$ are vectors containing
these matrices. Hence, the scalar product of these vectors is given by
\begin{equation}
\boldsymbol{I}\cdot\boldsymbol{S}_{\mathrm{h}}=\sum_{j=1}^{3}\boldsymbol{I}_{j}\boldsymbol{S}_{\mathrm{h}j}.
\end{equation}
The components of the matrices $\boldsymbol{I}_{i}$ read~\cite{25,100}
\begin{equation}
I_{i,\,jk}=-i\hbar\varepsilon_{ijk}\label{eq:I}
\end{equation}
with the Levi-Civita symbol $\varepsilon_{ijk}$.

\textcolor{black}{We have to note that the matrices $\boldsymbol{I}_{i}$ 
of the quasi-spin $I=1$ given by Eq.~(\ref{eq:I})
are not the standard spin matrices $\boldsymbol{S}_{i}$ 
of spin one~\cite{Messiah2}. However, a unitary transformation 
can be found so that $\boldsymbol{U}^{\dagger}
\boldsymbol{I}_{i}\boldsymbol{U}=\boldsymbol{S}_{i}$ holds.
The corresponding transformation matrix reads
\begin{equation}
\boldsymbol{U}=\frac{1}{\sqrt{2}}\left(\begin{array}{ccc}
-1 & 0 & 1\\
-i & 0 & -i\\
0 & \sqrt{2} & 0
\end{array}\right).
\end{equation}
Since in Ref.~\cite{Messiah2} the behavior of the standard spin
matrices under symmetry operations such as
time reversal and reflections are given, 
we will use the standard spin matrices 
in the following but denote them also by $\boldsymbol{I}_{i}$.}

The spin-orbit coupling $H_{\mathrm{so}}$ in Eq.~(\ref{eq:Hh}) is given by~\cite{25,7}
\begin{equation}
H_{\mathrm{so}}=\frac{2}{3}\Delta\left(1+\frac{1}{\hbar^{2}}\boldsymbol{I}\cdot\boldsymbol{S}_{\mathrm{h}}\right)\label{eq:soc}
\end{equation}
with the spin-orbit coupling constant $\Delta$. 
This coupling can be diagonalized by introducing the effective
hole spin $J=I+S_{\mathrm{h}}$. We choose the form of the
spin-orbit coupling so that the energy of the valence band with $J=1/2$ remains unchanged
while the two valence bands with $J=3/2$ are shifted by an amount of $\Delta$ towards
lower energies.
Note that we neglect the central-cell corrections treated in Ref.~\cite{200} in the Hamiltonian
as they do not affect the exciton states of high energy considered here. 

\textcolor{black}{The expression for $H_{\mathrm{h}}\left(\boldsymbol{p}_{\mathrm{h}}\right)$
can be written in terms of irreducible tensors (see, e.g., Refs.~\cite{ED,7_11,100,125}):
\begin{widetext}
\begin{eqnarray}
H_{\mathrm{h}}\left(\boldsymbol{p}_{\mathrm{h}}\right) & = & H_{\mathrm{so}}+\frac{\gamma_{1}}{2m_{0}}p_{\mathrm{h}}^{2}+\frac{\gamma'_{1}}{2\hbar^{2}m_{0}}\left[-\frac{\mu'}{3}P_{\mathrm{h}}^{(2)}\cdot I^{(2)}+\frac{\delta'}{3}\left(\sum_{k=\pm4}\left[P_{\mathrm{h}}^{(2)}\times I^{(2)}\right]_{k}^{(4)}+\frac{\sqrt{70}}{5}\left[P_{\mathrm{h}}^{(2)}\times I^{(2)}\right]_{0}^{(4)}\right)\right]
\end{eqnarray}
\end{widetext}
In this case one can clearly distinguish between the terms having spherical symmetry
and the terms having cubic symmetry.
While the first three terms have spherical symmetry, the last part with the coefficient
$\delta'$ has cubic symmetry.
The coefficients $\mu'$ and $\delta'$
are given in terms of the three Luttinger parameters as
$\mu'=\left(6\gamma_{3}+4\gamma_{2}\right)/5\gamma'_{1}$ and
$\delta'=\left(\gamma_{3}-\gamma_{2}\right)/\gamma'_{1}$
with $\gamma'_{1}=\gamma_{1}+m_{0}/m_{\mathrm{e}}$~\cite{7_11,7,100}.}

When applying external fields, the
corresponding Hamiltonian is obtained via the minimal substitution. 
We additionally introduce relative and center of mass coordinates~\cite{90,AM,91}.
Hence, we replace the coordinates and momenta of electron and hole with
\begin{subequations}
\begin{eqnarray}
\boldsymbol{r}_{\mathrm{e}} & = & \boldsymbol{R}+\left(m_{\mathrm{h}}/M\right)\boldsymbol{r},\\
\boldsymbol{r}_{\mathrm{h}} & = & \boldsymbol{R}-\left(m_{\mathrm{e}}/M\right)\boldsymbol{r},\\
\boldsymbol{p}_{\mathrm{e}} & = & \left(m_{\mathrm{e}}/M\right)\boldsymbol{P}+\boldsymbol{p}+e\boldsymbol{A}\left(\boldsymbol{r}\right),\\
\boldsymbol{p}_{\mathrm{h}} & = & \left(m_{\mathrm{h}}/M\right)\boldsymbol{P}-\boldsymbol{p}+e\boldsymbol{A}\left(\boldsymbol{r}\right),
\end{eqnarray}
\end{subequations}
where $M=m_{\mathrm{e}}+m_{\mathrm{h}}=m_{\mathrm{e}}+m_{0}/\gamma_1$ \textcolor{black}{denotes the yellow exciton mass.}
Then the Hamiltonian of the exciton reads~\cite{34,33,39,TOE,44,90,91}
\begin{eqnarray}
H_{\mathrm{exc}} & = & E_{\mathrm{g}} + V\left(\boldsymbol{r}\right)+e\Phi\left(\boldsymbol{r}\right)+H_{B}\nonumber \\
 & + & H_{\mathrm{e}}\left(\left(m_{\mathrm{e}}/M\right)\boldsymbol{P}+\boldsymbol{p}+e\boldsymbol{A}\left(\boldsymbol{r}\right)\right)\nonumber \\
 & + & H_{\mathrm{h}}\left(\left(m_{\mathrm{h}}/M\right)\boldsymbol{P}-\boldsymbol{p}+e\boldsymbol{A}\left(\boldsymbol{r}\right)\right).
 \label{eq:H}
\end{eqnarray}
We use the vector potential $\boldsymbol{A}=\left(\boldsymbol{B}\times\boldsymbol{r}\right)/2$
of a constant magnetic field $\boldsymbol{B}$ and the electrostatic potential
$\Phi\left(\boldsymbol{r}\right)=-\boldsymbol{F}\cdot\boldsymbol{r}$ of a constant electric field $\boldsymbol{F}$.

Since the Hamiltonian depends only 
on the relative coordinate $\boldsymbol{r}$,
the generalized momentum of the center of mass is 
a good quantum number, i.e., $\left[\boldsymbol{P},\,H_{\mathrm{exc}}\right]=\boldsymbol{0}$,
and one can generally set $\boldsymbol{P}=\hbar\boldsymbol{K}$~\cite{24,AM,17_15}.
\textcolor{black}{When neglecting the exciton-phonon interaction, one can especially assume $\boldsymbol{K}\approx\boldsymbol{0}$,
as the wave vector of photons, by which the excitons are created, is very close
to the origin of the Brillouin zone~\cite{TOE}.}

The additional term $H_{B}$ in Eq.~(\ref{eq:H}) describes the energy
of the spins in the magnetic field~\cite{25,44,44_12,33}:
\begin{equation}
H_{B}=\mu_{\mathrm{B}}\left[g_{c}\boldsymbol{S}_{\mathrm{e}}+\left(3\kappa+g_{s}/2\right)\boldsymbol{I}-g_{s}\boldsymbol{S}_{\mathrm{h}}\right]\cdot\boldsymbol{B}/\hbar.
\end{equation}
Here $\mu_{B}$ denotes the Bohr magneton, $g_{s}\approx2$ the $g$-factor
of the hole spin $S_{\mathrm{h}}$, $g_{c}$ the $g$-factor of the
conduction band or the electron spin $S_{\mathrm{e}}$, and $\kappa$
the fourth Luttinger parameter. 
All relevant material parameters of $\mathrm{Cu_{2}O}$ are listed in Table~\ref{tab:1}.

\begin{table}

\protect\caption{Material parameters of $\mathrm{Cu_{2}O}$.\label{tab:1}}

\begin{centering}
\begin{tabular}{lll}
\hline 
band gap energy & $E_{\mathrm{g}}=2.17208\,\mathrm{eV}$ & \cite{GRE}\tabularnewline
electron mass & $m_{\mathrm{e}}=0.99\, m_{0}$ & \cite{M2}\tabularnewline
dielectric constant & $\varepsilon=7.5$ & \cite{SOK1_82L1}\tabularnewline
spin-orbit coupling & $\Delta=0.131\,\mathrm{eV}$ & \cite{80}\tabularnewline
Luttinger parameters & $\gamma_{1}=1.76$ & \cite{80,100}\tabularnewline
 & $\gamma_2=0.7532$ & \cite{80,100}\tabularnewline
 & $\gamma_3=-0.3668$ & \cite{80,100}\tabularnewline
 & $\kappa=-0.5$ & \cite{125}\tabularnewline
$g$-factor of cond. band & $g_{\mathrm{c}}=2.1$ & \cite{29a}\tabularnewline
\hline 
\end{tabular}
\par\end{centering}

\end{table}

As we will show in Sec.~\ref{sec:analytical}, the symmetry
breaking in the system depends on the orientation
of the fields with respect to the crystal lattice.
We will denote the orientation of $\boldsymbol{B}$ and $\boldsymbol{F}$
in spherical coordinates via
\begin{equation}
\boldsymbol{B}\left(\varphi,\,\vartheta\right)=B\left(\begin{array}{c}
\cos\varphi\sin\vartheta,\\
\sin\varphi\sin\vartheta\\
\cos\vartheta
\end{array}\right)\label{eq:spherical_coord}
\end{equation}
and similar for $\boldsymbol{F}$ in what follows.

Before we solve the Schr\"odinger equation corresponding to the Hamiltonian~(\ref{eq:H}),
we rotate the coordinate system to make
the quantization axis coincide with the direction of the magnetic field (see Appendix~\ref{sub:Hamiltonianrm})
and then express the Hamiltonian~(\ref{eq:H}) in terms of irreducible tensors~\cite{ED,7_11,44}.

\subsection{Complete basis\label{sub:Complete basis}}

For our numerical investigations, we calculate a matrix representation of
the Schr\"odinger equation corresponding to the Hamiltonian 
$H_{\mathrm{exc}}$ of Eq.~(\ref{eq:H}) using a complete basis. 

As regards the angular momentum part of the basis,
we have to consider 
that the spin orbit coupling $H_{\mathrm{so}}$
couples the quasispin $I$ and the hole spin $S_{\mathrm{h}}$ to the 
effective hole spin $J=I+S_{\mathrm{h}}$. The
remaining parts of the kinetic energy of the hole
couple the effective hole spin $J$ and the angular momentum $L$ of the exciton
to the effective angular momentum $F=L+J$.
The electron spin $S_{\mathrm{e}}$ or its $z$ component $M_{S_{\mathrm{e}}}$
is a good quantum number.
For the radial
part of the exciton wave function we use the Coulomb-Sturmian functions of Ref.~\cite{S1}
\begin{equation}
U_{NL}\left(r\right)=N_{NL}\left(2\rho\right)^{L}e^{-\rho}L_{N}^{2L+1}\left(2\rho\right)\label{eq:U}
\end{equation}
with $\rho=r/\alpha$, a normalization factor $N_{NL}$,
the associated Laguerre polynomials $L_{n}^{m}\left(x\right)$ and
an arbitrary scaling parameter $\alpha$. 
Note that we use the radial quantum number
$N$, which is related to the principal quantum number $n$ via $n=N+L+1$.
Finally, we make the following ansatz for the exciton wave function
\begin{subequations}
\begin{eqnarray}
\left|\Psi\right\rangle  & = & \sum_{NLJFM_{F}}c_{NLJFM_{F}}\left|\Pi\right\rangle\left|S_{\mathrm{e}}, M_{S_{\mathrm{e}}}\right\rangle,\\
\nonumber \\
\left|\Pi\right\rangle  & = & \left|N,\, L;\,\left(I,\, S_{\mathrm{h}}\right)\, J;\, F,\, M_{F}\right\rangle\label{eq:basis}
\end{eqnarray}\label{eq:ansatz}%
\end{subequations}
with complex coefficients $c$. 
The parenthesis and semicolons in
Eq.~(\ref{eq:basis}) shall illustrate the coupling scheme of the
spins and the angular momenta.

Inserting the ansatz~(\ref{eq:ansatz}) in the Schr\"odinger
equation $H\Psi=E\Psi$ 
and multiplying from the left with
another basis state $\left\langle \Pi'\right|$
yields a matrix representation 
of the Schr\"odinger equation of the form~\cite{175}
\begin{equation}
\boldsymbol{D}\boldsymbol{c}=E\boldsymbol{M}\boldsymbol{c}.\label{eq:gev}
\end{equation}
The vector $\boldsymbol{c}$ contains the coefficients of the expansion~(\ref{eq:ansatz}).
Since the functions $U_{NL}\left(r\right)$
actually depend on the coordinate $\rho=r/\alpha$, we substitute
$r\rightarrow\rho\alpha$ in the Hamiltonian~(\ref{eq:H}) and multiply
the corresponding Schr\"odinger equation by $\alpha^{2}$.
All matrix elements which enter the hermitian matrices $\boldsymbol{D}$ and
$\boldsymbol{M}$ can be calculated similarly to the 
matrix elements given in Refs.~\cite{100,125}.
The generalized eigenvalue problem~(\ref{eq:gev})
is finally solved using an appropriate LAPACK routine~\cite{Lapack}.

Since in numerical calculations the basis cannot be infinitely large, 
the values of the quantum numbers
are chosen in the following way: For each value of $n=N+L+1\leq n_{\mathrm{max}}$ we use
\begin{eqnarray}
L & = & 0,\,\ldots,\, n-1,\nonumber \\
J & = & 1/2,\,3/2,\nonumber \\
F & = & \left|L-J\right|,\,\ldots,\,\min\left(L+J,\, F_{\mathrm{max}}\right),\\
M_{F} & = & -F,\,\ldots,\, F.\nonumber 
\end{eqnarray}
The values $F_{\mathrm{max}}$ and $n_{\mathrm{max}}$ are chosen
appropriately large so that as many eigenvalues as possible converge. Additionally,
we can use the scaling parameter $\alpha$ to enhance convergence~\cite{S1}.
However, it should be noted that the value of $\alpha$ does not influence the theoretical
results for the exciton energies in any way, i.e., the converged
results do not depend on the value of $\alpha$.

\subsection{Level spacing distributions\label{sub:Level-spacing-distributions}}

%
%

Having solved the generalized eigenvalue problem~(\ref{eq:gev})
the level statistics of the exciton spectra can be determined.
Before analyzing the nearest-neighbor spacings,
we have to unfold the spectra to obtain 
a constant mean spacing~\cite{225,GUE1,QSC,QC_1,QC_16}.
The unfolding procedure separates the average behavior
of the non-universal spectral density from 
universal spectral fluctuations
and yields a spectrum in which the mean level spacing
is equal to unity~\cite{GUE4}.
We leave out
a certain number of low-lying sparse levels to remove individual but
nontypical fluctuations~\cite{GUE1}.

Since the external fields
break all symmetries in the system and limit the
convergence of the solutions of the 
generalized eigenvalue problem with high energies~\cite{100},
the number of level spacings analyzed here
is comparatively small.
In this case, the cumulative distribution function~\cite{GUE2}
\begin{equation}
F(s)=\int_{0}^{s}P(x)\,\mathrm{d}x
\end{equation}
is often more meaningful than
histograms of the level
spacing probability distribution function $P(s)$.

We will compare our results with the
distribution functions known from random
matrix theory~\cite{QC_1,QC}:
the Poissonian distribution 
\begin{equation}
P_{\mathrm{P}}(s)=e^{-s}
\end{equation}
for non-interacting energy levels, the Wigner distribution 
\begin{equation}
P_{\mathrm{GOE}}(s)=\frac{\pi}{2}\,se^{-\pi s^2/4},\label{eq:GOE}
\end{equation} 
and the distribution 
\begin{equation}
P_{\mathrm{GUE}}(s)=\frac{32}{\pi^2}\,s^2e^{-4 s^2/\pi}\label{eq:GUE}
\end{equation}
for systems without any antiunitary symmetry.
Note that the most characteristic feature of GOE or GUE statistics is
the linear or quadratic level repulsion for small $s$, respectively.

In Ref.~\cite{GUE4} also analytical expressions for the spacing distribution
functions in the transition region between the different statistics
have been derived using random matrix theory for $2\times 2$ matrices.
As in our case only the transition from GOE to GUE statistics
will be important, we only give the analytical formula for this transition:
\begin{subequations}
\begin{equation}
P_{\mathrm{GOE}\rightarrow\mathrm{GUE}}\left(s;\,\lambda\right) = Cse^{-D^{2}s^{2}}\mathrm{erf}\left(\frac{Ds}{\lambda}\right)
\end{equation}
with
\begin{eqnarray}
D\left(\lambda\right) & = & \frac{\sqrt{1+\lambda^{2}}}{\sqrt{\pi}}\left(\frac{\lambda}{1+\lambda^{2}}+\mathrm{arccot}\left(\lambda\right)\right),\\
\nonumber \\
C\left(\lambda\right) & = & 2\sqrt{1+\lambda^{2}}D\left(\lambda\right)^{2}.
\end{eqnarray}\label{eq:GOEGUE}%
\end{subequations}
For the special cases of $\lambda\rightarrow 0$ or $\lambda\rightarrow\infty$
GOE or GUE statistics is obtained, respectively.
However, already for $\lambda\gtrsim 0.8$ the transition to
GUE statistics is almost completed~\cite{GUE4}.

As in Ref.~\cite{GUE4}, we calculate the distribution functions
for $\lambda=0.01\times 1000^{(k-1)/999}$ with $k=1,\ldots,1000$ and then
numerically integrate the results to obtain the corresponding 
cumulative distribution functions
$F_{\mathrm{GOE}\rightarrow\mathrm{GUE}}\left(s;\,\lambda\right)$.

\section{Fields not oriented in symmetry plane of the lattice\label{sec:analytical}}

In a previous paper~\cite{175} we have shown analytically
that the last remaining antiunitary symmetry known
from the hydrogen atom in external fields is broken
for the exciton Hamiltonian~(\ref{eq:H}) if the plane spanned by the
external fields is not identical to one of the symmetry planes of the solid.
Here we discuss this symmetry breaking in more detail
and also explain that the presence of quasi-particle interactions
will not restore the broken symmetries.

In the special case of $\gamma_2=\gamma_3=0$, the exciton
Hamiltonian~(\ref{eq:H}) is of the same form as the 
Hamiltonian of a hydrogen atom in external fields. 
It is well known that this Hamiltonian
is invariant under the combined symmetry 
of time inversion $K$ followed by a reflection $S_{\hat{\boldsymbol{n}}}$ 
at the specific plane spanned by both fields~\cite{QSC}.
This plane is given by the normal vector 
\begin{equation}
\hat{\boldsymbol{n}}=\left(\boldsymbol{B}\times\boldsymbol{F}\right)/\left|\boldsymbol{B}\times\boldsymbol{F}\right|\label{eq:nvec}
\end{equation}
or $\hat{\boldsymbol{n}}\perp\hat{\boldsymbol{B}}=\boldsymbol{B}/B$ if $\boldsymbol{F}=\boldsymbol{0}$ holds.
Due to this antiunitary symmetry, the hydrogen-like system 
shows GOE statistics in the chaotic regime~\cite{GUE9,GUE1}. However, we have to note
that this is the last remaining antiunitary symmetry when applying external
fields.

Since the hydrogen atom is spherically symmetric in the field-free case,
it makes no difference whether the magnetic field is
oriented in $z$ direction or not. 
However, in a semiconductor with $\delta'\neq 0$ the exciton Hamiltonian
has cubic symmetry and the orientation
of the external fields with respect to the crystal axis
of the lattice becomes important.
Any rotation of the coordinate system with the aim 
of making the $z$ axis coincide with the direction of the magnetic
field will also rotate the cubic crystal lattice.
Hence, the antiunitary symmetry
mentioned above is only present if the plane spanned by both fields
is identical to one of the nine symmetry planes of the cubic lattice
since then the reflection $S_{\hat{\boldsymbol{n}}}$
transforms the lattice into itself.
\textcolor{black}{However, if none of the 
normal vectors $\hat{\boldsymbol{n}}_i$ of these 
nine symmetry planes (cf.~Appendix~\ref{sec:normalvec})}
is parallel to the direction $\hat{\boldsymbol{n}}$ given in Eq.~(\ref{eq:nvec}),
or, in the case of $\boldsymbol{F}=\boldsymbol{0}$,
if none of these vectors is perpendicular to
$\hat{\boldsymbol{B}}=\boldsymbol{B}/B$, the last antiunitary symmetry is broken.
In these cases the commutator of the exciton Hamiltonian~(\ref{eq:H}) 
with the operator $KS_{\hat{\boldsymbol{n}}}$
does not vanish as we will show in the following.

Under time inversion $K$ and reflections $S_{\hat{\boldsymbol{n}}}$
at a plane perpendicular to a normal vector $\hat{\boldsymbol{n}}$
the vectors of position $\boldsymbol{r}$, momentum $\boldsymbol{p}$
and spin $\boldsymbol{S}$ transform according to~\cite{Messiah2}
\begin{subequations}
\begin{eqnarray}
K\boldsymbol{r}K^{\dagger}& = &\boldsymbol{r},\\
K\boldsymbol{p}K^{\dagger}& = &-\boldsymbol{p},\\
K\boldsymbol{S}K^{\dagger}& = &-\boldsymbol{S},
\end{eqnarray}
\end{subequations}
and
\begin{subequations}
\begin{eqnarray}
S_{\hat{\boldsymbol{n}}}\boldsymbol{r}S_{\hat{\boldsymbol{n}}}^{\dagger} & = & \boldsymbol{r}-2\hat{\boldsymbol{n}}\left(\hat{\boldsymbol{n}}\cdot\boldsymbol{r}\right),\\
S_{\hat{\boldsymbol{n}}}\boldsymbol{p}S_{\hat{\boldsymbol{n}}}^{\dagger} & = & \boldsymbol{p}-2\hat{\boldsymbol{n}}\left(\hat{\boldsymbol{n}}\cdot\boldsymbol{p}\right),\\
S_{\hat{\boldsymbol{n}}}\boldsymbol{S}S_{\hat{\boldsymbol{n}}}^{\dagger} & = & -\boldsymbol{S}+2\hat{\boldsymbol{n}}\left(\hat{\boldsymbol{n}}\cdot\boldsymbol{S}\right).
\end{eqnarray}
\end{subequations}

For all orientations of the external fields the hydrogen-like part of the Hamiltonian~(\ref{eq:H}) is
invariant under $KS_{\hat{\boldsymbol{n}}}$ with $\hat{\boldsymbol{n}}$ given by Eq.~(\ref{eq:nvec}). However, other parts
of the Hamiltonian such as $H_c=\left(p_{1}^{2}\boldsymbol{I}_{1}^{2}+\mathrm{c.p.}\right)$ 
are not invariant if the fields are not oriented in one symmetry plane of the lattice. 
For example, for the case with $\boldsymbol{B}\left(0,\,0\right)$ and $\boldsymbol{F}\left(\pi/6,\,\pi/2\right)$,
we obtain 
\begin{eqnarray}
& & S_{\hat{\boldsymbol{n}}}KH_{c}K^{\dagger}S_{\hat{\boldsymbol{n}}}^{\dagger}-H_{c}\nonumber\\
& = & 1/8\left[2\sqrt{3}\left(\boldsymbol{I}_{2}^{2}-\boldsymbol{I}_{1}^{2}\right)p_{1}p_{2}\right.\nonumber\\
& + & 3\left(\boldsymbol{I}_{1}^{2}p_{2}^{2}+\boldsymbol{I}_{2}^{2}p_{1}^{2}\right)-3\left(\boldsymbol{I}_{1}^{2}p_{1}^{2}+\boldsymbol{I}_{2}^{2}p_{2}^{2}\right)\nonumber\\
& + & \left.\left\{\boldsymbol{I}_{1},\boldsymbol{I}_{2}\right\}\left(2\sqrt{3}\left(p_{2}^{2}-p_{1}^{2}\right)+12p_{1}p_{2}\right)\right]\neq 0\label{eq:SKHCKS}
\end{eqnarray}
with $\hat{\boldsymbol{n}}=\left(-1/2,\,\sqrt{3}/2,\,0\right)^{\mathrm{T}}$.
Note that even though $H_c$ does not depend on the external fields,
the normal vector $\hat{\boldsymbol{n}}$ is determined 
by these fields via Eq.~(\ref{eq:nvec}). Otherwise, the
hydrogen-like part of the Hamiltonian would not be invariant under 
$KS_{\hat{\boldsymbol{n}}}$.

\begin{figure*}[t]
\begin{centering}
\includegraphics[width=2.0\columnwidth]{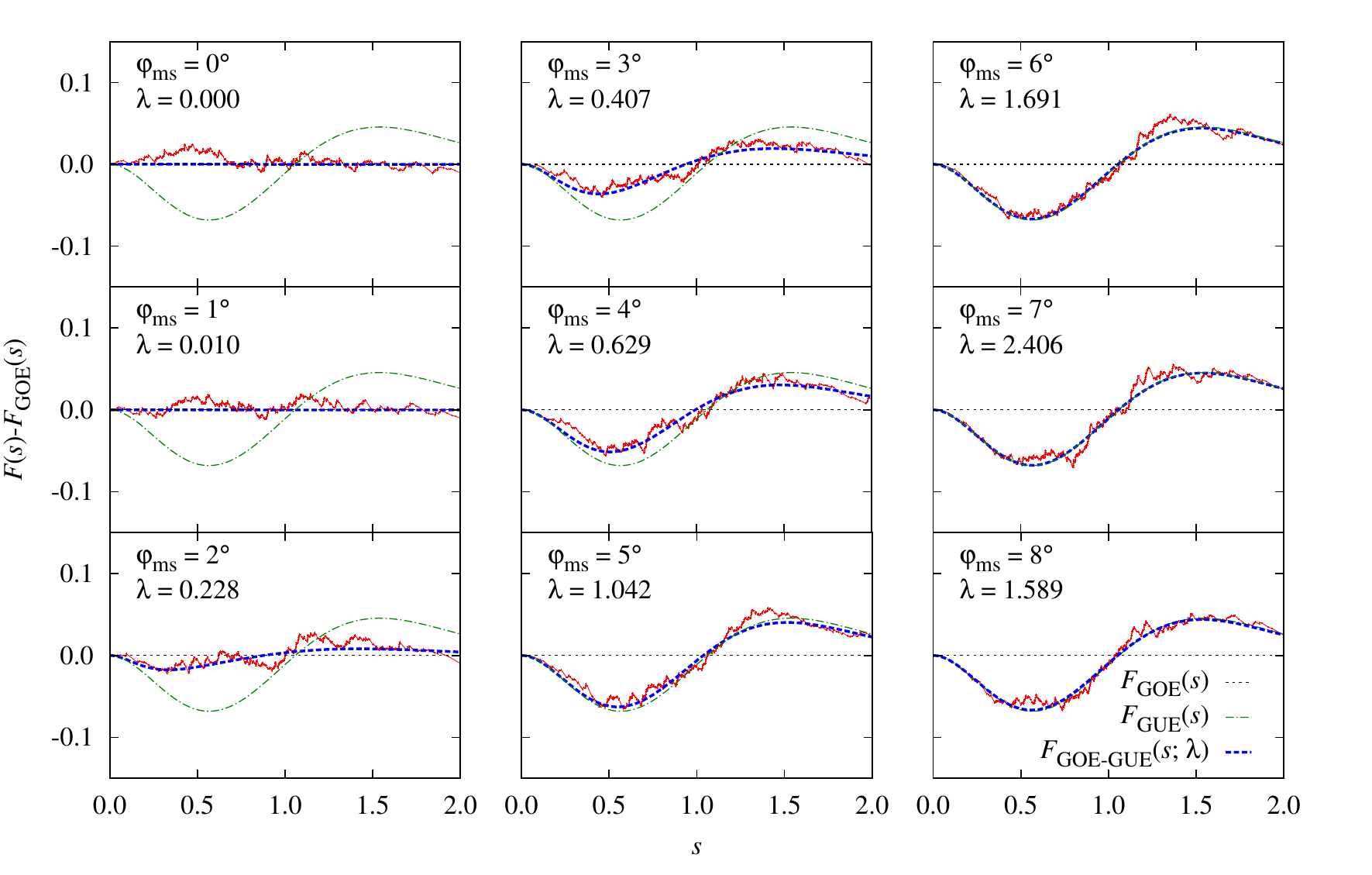}
\par\end{centering}

\protect\caption{Transition from GOE to GUE statistics
when deflecting the field $\boldsymbol{F}_{\mathrm{ms}}$ in Eq.~(\ref{eq:Fexamp}) from the symmetry plane
$y=0$ by an angle $\varphi_{\mathrm{ms}}$. 
For the magnetic field we have set $\boldsymbol{B}\left(\varphi=0,\,\vartheta=\pi/6\right)$
with $B=3\,\mathrm{T}$. To obtain enough eigenvalues for a statistical
evaluation, we used the simplified model of Ref.~\cite{175}, in which the spins
of the electron and hole are neglected.
To visualize the differences between the cumulative distribution functions
more clearly, we subtract $F_{\mathrm{GOE}}(s)$ from them.
The data points (red) were fitted with the analytical
function $F_{\mathrm{GOE}\rightarrow\mathrm{GUE}}\left(s;\,\lambda\right)$ defined in Sec.~\ref{sub:Level-spacing-distributions}.
The optimum values of the fit parameter $\lambda$ are given in each panel and are also shown in Fig.~\ref{fig:phi2}.
One can observe a good agreement between the numerical data and the 
analytical function describing the transition between the two statistics in dependence on $\lambda$.
For further information see text.~\label{fig:numer}}
\end{figure*}

\begin{figure}[t]
\begin{centering}
\includegraphics[width=1.0\columnwidth]{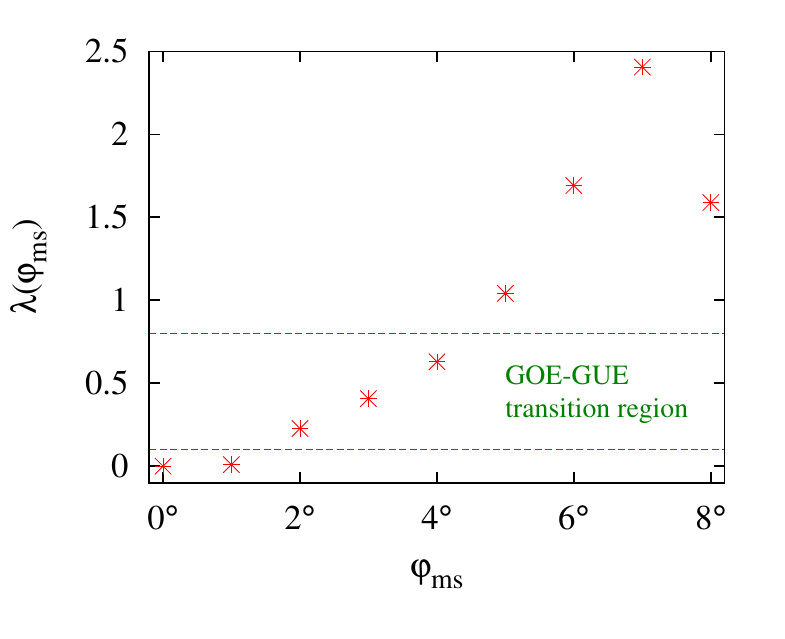}
\par\end{centering}

\protect\caption{Optimum values of the fit parameter $\lambda$
in dependence on the angle $\varphi_{\mathrm{ms}}$ for the situation presented in Fig.~\ref{fig:numer}.
One can see that the value of $\lambda$ increases very rapidly
with increasing $\varphi_{\mathrm{ms}}$. Already for $\varphi_{\mathrm{ms}}=5\degree$ the transition to
GUE statistics is completed. As regards the value of $\lambda$ for 
$\varphi_{\mathrm{ms}}=8\degree$, we have to note that the function 
$F_{\mathrm{GOE}\rightarrow\mathrm{GUE}}\left(s;\,\lambda\right)$
only slightly varies for $\lambda\geq 0.8$ and hence small fluctuations
in the numerical results will lead to a strong change in $\lambda$. For the
transition between GOE and GUE statistics only the range of $0.1\leq\lambda\leq 0.8$
is of importance (green dahed lines) (cf.~Ref.~cite{225}). For $\varphi_{\mathrm{ms}}>8\degree$ 
it is always $\lambda>0.8$ until $\varphi_{\mathrm{ms}}\approx 176\degree$ [cf.~Eq.~(\ref{eq:nBFms})].
\label{fig:phi2}}

\end{figure}

\begin{figure}
\begin{centering}
\includegraphics[width=1.0\columnwidth]{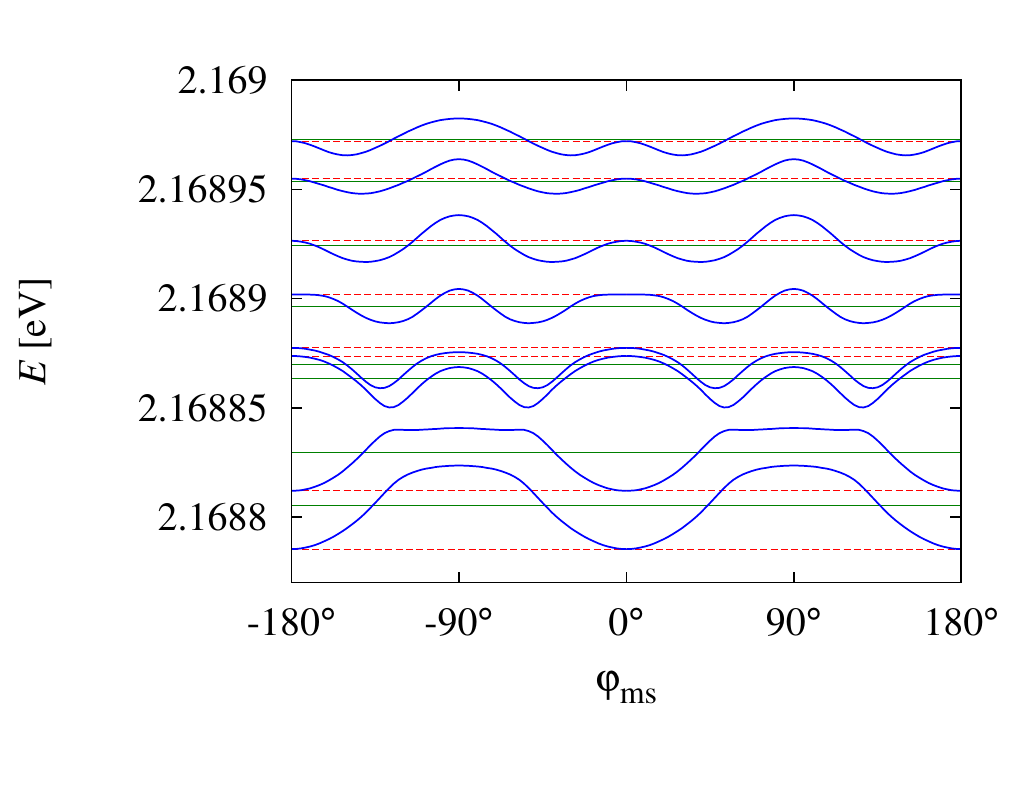}
\par\end{centering}

\protect\caption{Dependence of the energy of specific exciton states
on the angle $\varphi_{\mathrm{ms}}$ of the field 
$\boldsymbol{F}_{\mathrm{ms}}$. For the magnetic field we have 
set $\boldsymbol{B}\left(\varphi=0,\,\vartheta=\pi/6\right)$
with $B=3\,\mathrm{T}$. 
It can be seen that for $\varphi_{\mathrm{ms}}$
and $\pi+\varphi_{\mathrm{ms}}$ the exciton energies (blue solid lines) are shifted in the same direction
with respect to the energy at $\varphi_{\mathrm{ms}}=0$ (red dashed lines).
The average energy (green solid lines) often clearly differs from the energy at $\varphi_{\mathrm{ms}}=0$.
\label{fig:phi3}}

\end{figure}

\begin{figure}
\begin{centering}
\includegraphics[width=0.9\columnwidth]{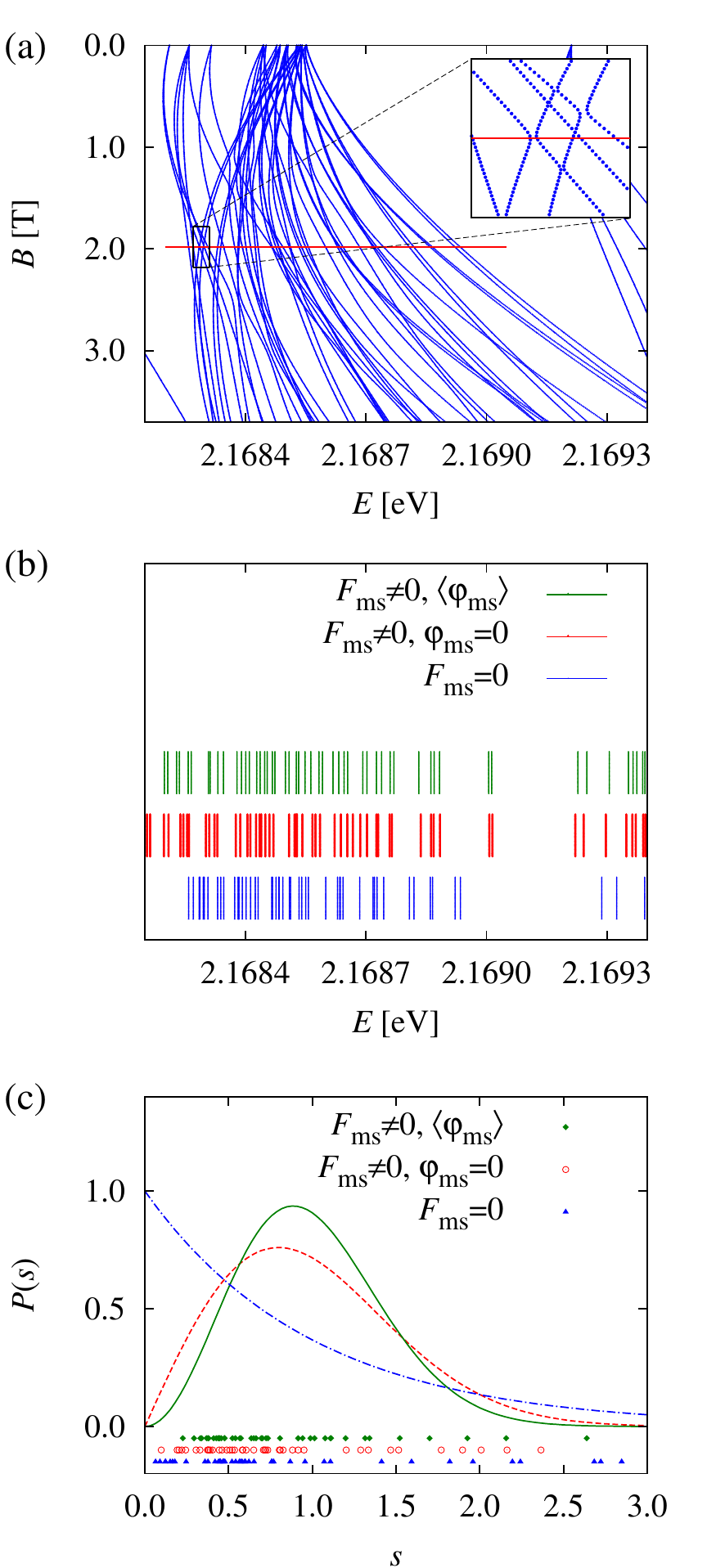}
\par\end{centering}

\protect\caption{(a) Splitting of the $n=5$ exciton states of $\mathrm{Cu_{2}O}$ 
in an external magnetic field $\boldsymbol{B}=\boldsymbol{B}\left(\varphi=0,\,\vartheta=\pi/6\right)$
with $F_{\mathrm{ms}}=0$. At $B\approx 1.98\,\mathrm{T}$ an
avoided crossing can be observed (see inset and red box).
(b) Energy of the $n=5$ states for $B=1.98\,\mathrm{T}$ and (i) $F_{\mathrm{ms}}=0$ (blue lines),
(ii) ${F}_{\mathrm{ms}}=9.57\times 10^{3}\,\mathrm{V/m}$ given by Eq.~(\ref{eq:Fexamp}) with $\varphi_{\mathrm{ms}}=0$ (red lines),
and (iii) ${F}_{\mathrm{ms}}$ given by Eq.~(\ref{eq:Fexamp}) but taking the position of the states 
when averaging over $\varphi_{\mathrm{ms}}=0$ (green lines).
(c) Normalized spacings for the three cases considered.
It can be seen that the motional Stark effect further suppresses small spacings.
For a comparison, we also show the distribution functions for Poisson statistics (blue dash-dotted line),
GOE statistics (red dashed line), and GUE statistics (green solid line).
\label{fig:full}}

\end{figure}

Since the expression in Eq.~(\ref{eq:SKHCKS})
is not equal to zero, we have shown for
$\boldsymbol{B}\left(0,\,0\right)$ and $\boldsymbol{F}\left(\pi/6,\,\pi/2\right)$
that the generalized time-reversal symmetry
of the hydrogen atom is broken
for excitons due to the cubic symmetry of the semiconductor.
The same calculation can also be performed for other orientations
of the external fields.
As we have stated above, the antiunitary symmetry remains
unbroken only for those specific orientations of the fields, where
$\hat{\boldsymbol{n}}$ given by Eq.~(\ref{eq:nvec}) is parallel to one of
the normal vectors in Eq.~(\ref{eq:ni}).

In the previous treatment we have neglected quasi-particle interactions
like the exciton-phonon interaction.
Hence, one may ask whether these interactions are able to 
restore the broken symmetries.

It is well known that when adding an additional interaction
to a Hamiltonian, this interaction will often further reduce
the symmetry of the problem and not increase it.
Indeed, it is not possible that the 
effects of the band structure and quasi-particle interactions
on the symmetry or the level spacing statistics will cancel 
each other out, in particular for all
values of the external field strengths. 
The quasi-particle interaction would have to have the same form as the
operators in our Hamiltonian to make the commutator of the Hamiltonian
and the symmetry operator $KS_{\hat{\boldsymbol{n}}}$ vanish.
However, if we, e.g., consider the interaction between excitons and phonons,
the interaction operators~\cite{SST} look quite different than the operators in
the exciton Hamiltonian~(\ref{eq:H}). 
Hence, phonons or other interactions in the solid
do not restore the broken antiunitary symmetries
if the external fields are not oriented in one symmetry plane of the solid.

\section{Fields oriented in symmetry plane of the lattice\label{sec:phonons}}

In this section we discuss the case that the plane spanned by
the external fields coincides with a symmetry plane of the lattice.
Without the exciton-phonon interaction one would expect to observe only GOE statistics
according to the explanations given in Sec.~\ref{sec:analytical}.
However, recent experiments indicate that the
spectrum of magnetoexcitons reveals GUE statistics for \emph{all} orientations
of the magnetic field applied~\cite{QC,QC2}.
Hence, we will now concentrate on the effects of the exciton-phonon
interaction in more detail and show that they lead, in combination
with the cubic valence band structure, to a breaking of all antiunitary symmetries
for an arbitrary orientation of the external fields.


\textcolor{black}{The Hamiltonian describing the exciton-phonon interaction~\cite{SST}
does not only depend on the relative coordinate $\boldsymbol{r}$ but also
on the coordinate of the center of mass $\boldsymbol{R}$.
Hence, when considering the Hamiltonian of excitons \emph{and} photons
the momentum of the center of mass $\boldsymbol{P}$
is not a good quantum number, i.e., the Hamiltonian and the
operator $\boldsymbol{P}$ no longer commute.}
Consequently, we are not allowed to set the momentum of
the center of mass to zero, as has been done in the calculation
of Sec.~\ref{sec:analytical}, but have to treat the complete problem.
However, the consideration of the valence band structure,
a finite momentum of the center of mass, the external fields, 
and the phonons is very complicated.
Hence, we concentrate only on the main effects 
to show that the exciton-phonon
interaction will lead to a breaking of all 
antiunitary symmetries even if the
plane spanned by the external fields is 
identical to a symmetry plane of the lattice.

When considering a finite momentum $\boldsymbol{P}=\hbar\boldsymbol{K}$ 
of the exciton center of mass in an external magnetic field, 
the motional Stark effect occurs~\cite{AM}.
Since the insertion of a finite momentum of the center of mass
in the complete Hamiltonian~(\ref{eq:H}) is quite laborious (cf.~Ref.~\cite{250}),
we treat only the leading term of the motional Stark effect, which
has the form~\cite{AM} 
\begin{equation}
H_{\mathrm{ms}}=\frac{\hbar e}{M}\left(\boldsymbol{K}\times\boldsymbol{B}\right)\cdot\boldsymbol{r}
\end{equation}
with the isotropic exciton mass $M=m_{\mathrm{e}}+m_{\mathrm{h}}=m_{\mathrm{e}}+m_0/\gamma_{1}$.
Note that this term has the same form as the electric field term in the
Hamiltonian~(\ref{eq:H}).
Hence, the effect of the motional Stark effect is the same as that of an
external electric field and we can introduce a total electric field
\begin{equation}
\boldsymbol{F}_{\mathrm{tot}}=\boldsymbol{F}+\boldsymbol{F}_{\mathrm{ms}}=
\boldsymbol{F}-\frac{\hbar}{M}\left(\boldsymbol{K}\times\boldsymbol{B}\right).
\end{equation}
One could now, in principle, do the same calculation as in Eq.~(\ref{eq:SKHCKS})
to show that the antiunitary symmetry known from the hydrogen atom
is broken if the plane spanned by $\boldsymbol{B}$ and $\boldsymbol{F}_{\mathrm{tot}}$
is not identical to one symmetry plane of the solid.
However, we have to consider the specific properties, i.e., 
the size and the orientation, of the
motional stark field $\boldsymbol{F}_{\mathrm{ms}}$ related
to the size and the orientation of $\boldsymbol{K}$.

The size of the momentum $\hbar\boldsymbol{K}$
is determined by the interaction between excitons and phonons.
Instead of considering the huge number of phonon degrees of freedom,
we assume a thermal distribution at a finite temperature $T$.
The direction of $\boldsymbol{K}$ is then evenly distributed 
over the solid angle and its average size is determined by
\textcolor{black}{\begin{equation}
\frac{3}{2}k_{\mathrm{B}}T=\frac{\hbar^2 K^2}{2M}\label{eq:K_Boltz}
\end{equation}
with the Boltzmann constant $k_{\mathrm{B}}$.
We assume for all of our calculations a temperature of 
$T=0.8\,\mathrm{K}$, which is even slightly
smaller than the temperature in experiments~\cite{GRE}.}
The relation~(\ref{eq:K_Boltz}) leads to a field strength of
\textcolor{black}{\begin{equation}
F_{\mathrm{ms}}=\sqrt{\frac{3k_{\mathrm{B}}T}{M}}\,B.\label{eq:Fms}
\end{equation}}
Note that the value of $K$ determined by Eq.~(\ref{eq:K_Boltz})
is of the same order of magnitude as the value estimated
via experimental group velocity measurements of the 
$1S$ ortho exciton~\cite{GRE_28,P3_107_12}.

We will now show that the motional Stark field $\boldsymbol{F}_{\mathrm{ms}}$ leads 
to GUE statistics if the external magnetic field $\boldsymbol{B}$ 
is oriented in one of the symmetry planes of the lattice
In the general case, the magnetic field then fulfils $\boldsymbol{B}\perp\hat{\boldsymbol{n}}_i$
with one of the nine normal vectors $\hat{\boldsymbol{n}}_i$ given in Eq.~(\ref{eq:ni}).
In our numerical example we choose the magnetic field
\begin{equation}
\boldsymbol{B}=\boldsymbol{B}\left(\varphi=0,\,\vartheta=\pi/6\right)=
\frac{B}{2}\left(\begin{array}{c}
1\\
0\\
\sqrt{3}
\end{array}\right)\perp\hat{\boldsymbol{n}}_2\label{eq:Bex}
\end{equation}
with a constant field strength of $B=3\,\mathrm{T}$.
The external electric field is set to $\boldsymbol{F}=\boldsymbol{0}$.
The motional Stark field is oriented perpendicular to $\boldsymbol{B}$.
Hence, we assume it for $\varphi_{\mathrm{ms}}$
to be oriented perpendicular to the magnetic field and to be initially lying
in the same symmetry plane $y=0$ of the lattice.
Then $\boldsymbol{F}_{\mathrm{ms}}$ is deflected from
this plane, i.e., the field is rotated by an angle $\varphi_{\mathrm{ms}}$
about the axis given by the magnetic field of Eq.~(\ref{eq:Bex}): 
\begin{equation}
\boldsymbol{F}_{\mathrm{ms}}\left(\varphi_{\mathrm{ms}}\right)=
\frac{F_{\mathrm{ms}}}{2}\left(\begin{array}{c}
\sqrt{3}\cos\varphi_{\mathrm{ms}}\\
2\sin\varphi_{\mathrm{ms}}\\
-\cos\varphi_{\mathrm{ms}}
\end{array}\right).\label{eq:Fexamp}
\end{equation}
Here $F_{\mathrm{ms}}$ is given by Eq.~(\ref{eq:Fms}) with $B=3\,\mathrm{T}$
and $T=1.2\,\mathrm{K}$.
According to the explanations given in Sec.~\ref{sec:analytical}, we
expect to obtain GOE statistics with our numerical results only for the cases $\varphi_{\mathrm{ms}}=0$
and $\varphi_{\mathrm{ms}}=\pi$, since
\begin{equation}
\hat{\boldsymbol{n}}=\left(\boldsymbol{B}\times\boldsymbol{F}\right)/\left|\boldsymbol{B}\times\boldsymbol{F}\right|=
\frac{1}{2}\left(\begin{array}{c}
-\sqrt{3}\sin\varphi_{\mathrm{ms}}\\
2\cos\varphi_{\mathrm{ms}}\\
\sin\varphi_{\mathrm{ms}}
\end{array}\right)\label{eq:nBFms}
\end{equation}
is parallel to $\hat{\boldsymbol{n}}_2$ only for these two values of $\varphi_{\mathrm{ms}}$.
The decisive question is how fast the transition from GOE to GUE statistics
takes place if the field $\boldsymbol{F}_{\mathrm{ms}}$ is deflected from the
symmetry plane $y=0$.
This is shown in Fig.~\ref{fig:numer}.

As we have already stated in Ref.~\cite{175} and Sec.~\ref{sub:Level-spacing-distributions}, 
the number of eigenvalues which can be used for a statistical analysis is limited due
to the required computer memory or the limited size of our basis.
Therefore, to enhance the number of converged states, we used for the calculation of 
Fig.~\ref{fig:numer} the simplified model of Ref.~\cite{175} with $\Delta=H_B=0$,
$m_{\mathrm{e}}=m_0$, $\gamma_1=2$ and $\delta'=-0.15$. However, we expect a
qualitatively similar behavior for $\mathrm{Cu_{2}O}$, i.e., when considering $\Delta\neq 0$,
as we will discuss and show below. 

For a quantitative analysis the results are fitted with the 
function $F_{\mathrm{GOE}\rightarrow\mathrm{GUE}}\left(s;\,\lambda\right)$ 
[cf.~Eq.~(\ref{eq:GOEGUE})] describing the transition between both statistics.
We show the resulting values of the fit parameter $\lambda$ in Fig.~\ref{fig:phi2}.
It can be seen that the parameter $\lambda$ increases very rapidly with increasing
values of $\varphi_{\mathrm{ms}}$. Already for $\varphi_{\mathrm{ms}}=5\degree$
the statistics is almost purely GUE statistics. Hence, the motional Stark field
has a strong influence on the level spacing statistics.
This implies that for a majority of the orientations of $\boldsymbol{F}_{\mathrm{ms}}$
GUE statistics will be observable.
Our main argument for the observed level statistics is 
now that since the momentum $\boldsymbol{K}$ and hence also the field 
$\boldsymbol{F}_{\mathrm{ms}}$ is evenly distributed over the angle $\varphi_{\mathrm{ms}}$,
the exciton spectrum will show GUE statistics on average.

One might argue whether the effects of $\boldsymbol{F}_{\mathrm{ms}}$
cancel each other out if the field is evenly distributed over the solid angle.
This can be ruled out when considering the effect of the
field on the exciton states for all values of the angle $\varphi_{\mathrm{ms}}$
as shown for a selection of exciton states in Fig.~\ref{fig:phi3}.
It can be seen that the fields
$\boldsymbol{F}_{\mathrm{ms}}\left(\varphi_{\mathrm{ms}}\right)$
and $\boldsymbol{F}_{\mathrm{ms}}\left(\pi+\varphi_{\mathrm{ms}}\right)
=-\boldsymbol{F}_{\mathrm{ms}}\left(\varphi_{\mathrm{ms}}\right)$
shift the exciton states in the same direction and not
in opposite direction as regards their energies. Hence, on average
the exciton states are shifted towards higher or lower energies
and do not remain at their position.
This argument holds both when using the model with the
parameters of Ref.~\cite{175} \emph{and} when using all material parameters
of $\mathrm{Cu_{2}O}$. In Fig.~\ref{fig:phi3} the 
results for $\mathrm{Cu_{2}O}$ are shown.

Even though we cannot obtain enough converged exciton energies
for a statistical analysis when using the parameters of $\mathrm{Cu_{2}O}$,
we can use the small number of converged states to show
that the magneto Stark field has the small effect of increasing
level spacings, which is a characteristic feature of GUE compared to GOE statistics [cf.~Eqs.~(\ref{eq:GOE}) and~(\ref{eq:GUE})].

To this aim, we consider at first the spectrum
of $\mathrm{Cu_{2}O}$ in a magnetic field 
$\boldsymbol{B}\left(\varphi=0,\,\vartheta=\pi/6\right)$ 
to find an avoided crossing (see panel (a) of Fig.~\ref{fig:full}).
We then choose the magnetic field strength of $B=1.98\,\mathrm{T}$,
where an avoided crossing appears, to be fixed, and calculate the
spectrum in dependence on the angle $\varphi_{\mathrm{ms}}$.
The strength of the motional Stark field is given by Eq.~(\ref{eq:Fms}) with
$B=1.98\,\mathrm{T}$ and $T=1.2\,\mathrm{K}$.
We now calculate the energies of the states for the
following three cases, where the magnetic field strength is always
given by $B=1.98\,\mathrm{T}$: (i) $F_{\mathrm{ms}}=0$, 
(ii) $F_{\mathrm{ms}}=\sqrt{2k_{\mathrm{B}}T/M}\,B$ and $\varphi_{\mathrm{ms}}=0$,
(iii) $F_{\mathrm{ms}}=\sqrt{2k_{\mathrm{B}}T/M}\,B$ and taking the average
of the exciton energies over $\varphi_{\mathrm{ms}}$. These energies are shown 
in panel (b) of Fig.~\ref{fig:full}.
\textcolor{black}{We assume a constant density of states due
to the small energy range considered here. 
Then the normalized spacings between two 
neighboring exciton states are determined
as $s_i=(E_i-E_{i+1})/\bar{E}$ with $\bar{E}$
denoting the mean value of all spacings considered.}
One can see from panel (c) of Fig.~\ref{fig:full} that the
level spacings change for the three cases considered. 
Especially for small values of $s$ the spacing increases,
which illustrates the repulsion of levels and the transition to GUE statistics.

Overall, it can be stated that the exciton-phonon interaction
leads to a finite momentum of the center of mass
of the exciton, which is evenly distributed over the solid angle.
The size of this momentum is on average determined by the Boltzmann distribution.
In an external magnetic field this finite momentum causes
the motional Stark effect. The electric field corresponding to this effect
breaks in combination with the cubic lattice all antiunitary symmetries
in the system even if the plane spanned by the external fields
coincides with one symmetry plane of the lattice.

\section{Summary and outlook\label{sec:Summary}}

We have shown analytically that the combined presence of the
cubic valence band structure and external fields
breaks all antiunitary symmetries for excitons in $\mathrm{Cu_{2}O}$.
When neglecting the exciton-phonon interaction,
this symmetry breaking appears only if the plane spanned 
by the external fields is not identical to one of the
symmetry planes of the cubic lattice of $\mathrm{Cu_{2}O}$.
We have discussed that for these cases the additional presence
of the exciton-phonon interaction is not able to restore the
broken symmetries.

For the specific orientations of the external
fields, where the plane spanned by the fields
is identical to one of the
symmetry planes of the cubic lattice, the exciton-phonon
interaction becomes important.
This interaction
causes a finite momentum of the exciton center of mass, which
leads to the motional Stark effect in an external magnetic field.
If the cubic valence band structure is considered, the 
effective electric field connected with the motional Stark effect
finally leads to the breaking of all antiunitary symmetries.
Since the exciton-phonon interaction is always present in the solid,
we have thus shown that GUE statistics will be observable in all spectra 
of magnetoexcitons irrespective of the orientation of the external
magnetic field, which is in agreement with the 
experimental observations in Refs.~\cite{QC,QC2}.

\acknowledgments
F.S. is  grateful  for  support  from  the
Landesgraduiertenf\"orderung of the Land Baden-W\"urttemberg.

\appendix

\section{Hamiltonian \label{sub:Hamiltonianrm}}

Here we give the complete Hamiltonian of Eq.~(\ref{eq:H})
and describe the rotation necessary to make the quantization axis coincide 
with the direction of the magnetic field.
Let us write the Hamiltonian~(\ref{eq:H}) in the form
\begin{eqnarray}
H & = & H_0+(eB)H_1+(eB)^2 H_2 - e\boldsymbol{F}\cdot\boldsymbol{r}\nonumber \\
\nonumber \\
& + & E_{\mathrm{g}}-\frac{e^{2}}{4\pi\varepsilon_{0}\varepsilon}\frac{1}{r}+\frac{2}{3}\Delta\left(1+\frac{1}{\hbar^{2}}\boldsymbol{I}\cdot \boldsymbol{S}_{\mathrm{h}}\right)
\end{eqnarray}
with $B=\left|\boldsymbol{B}\right|$. Using
$\hat{B}_i=B_i/B$ with the components $B_i$ of $\boldsymbol{B}$, the terms
$H_0$, $H_1$, and $H_2$ are given by

\begin{widetext}
\begin{align}
H_{0} & = \: \frac{1}{2m_{0}}\left(\gamma_{1}'+4\gamma_{2}\right)\boldsymbol{p}^{2}+\frac{1}{\hbar^{2}m_{0}}\left(\eta_{1}+2\eta_{2}\right)\left(\boldsymbol{I}\cdot\boldsymbol{S}_{\mathrm{h}}\right)\boldsymbol{p}^{2}\nonumber \\
\displaybreak[2]\nonumber \\
& -\frac{3\gamma_{2}}{\hbar^{2}m_{0}}\left[\boldsymbol{I}_{1}^{2}p_{1}^{2}+\mathrm{c.p.}\right]-\frac{6\eta_{2}}{\hbar^{2}m_{0}}\left[\boldsymbol{I}_{1}\boldsymbol{S}_{\mathrm{h}1}p_{1}^{2}+\mathrm{c.p.}\right]\nonumber \\
\displaybreak[2]\nonumber \\
& -\frac{6\gamma_{3}}{\hbar^{2}m_{0}}\left[\left\{ \boldsymbol{I}_{1},\,\boldsymbol{I}_{2}\right\} p_{1}p_{2}+\mathrm{c.p.}\right]-\frac{6\eta_{3}}{\hbar^{2}m_{0}}\left[\left(\boldsymbol{I}_{1}\boldsymbol{S}_{\mathrm{h}2}+\boldsymbol{I}_{2}\boldsymbol{S}_{\mathrm{h}1}\right)p_{1}p_{2}+\mathrm{c.p.}\right],\label{eq:A2}
\\
\displaybreak[1]
\nonumber \\
H_{1} & = \: \frac{1}{2m_{0}}\left(\frac{2m_{0}}{m_{\mathrm{e}}}-\gamma_{1}'+4\gamma_{2}\right)\hat{\boldsymbol{B}}\cdot\boldsymbol{L}-\frac{1}{\hbar^{2}m_{0}}\left(\eta_{1}+2\eta_{2}\right)\left(\boldsymbol{I}\cdot\boldsymbol{S}_{\mathrm{h}}\right)\hat{\boldsymbol{B}}\cdot\boldsymbol{L}\nonumber \\
\displaybreak[2]\nonumber \\
& +\frac{1}{2m_{0}}\left[g_{c}\boldsymbol{S}_{\mathrm{e}}+\left(3\kappa+\frac{g_{s}}{2}\right)\boldsymbol{I}-g_{s}\boldsymbol{S}_{\mathrm{h}}\right]\cdot\hat{\boldsymbol{B}}\nonumber \\
\displaybreak[2]\nonumber \\
& +\frac{3\gamma_{2}}{\hbar^{2}m_{0}}\left[\boldsymbol{I}_{1}^{2}\left(\hat{B}_{2}r_{3}p_{1}-\hat{B}_{3}r_{2}p_{1}\right)+\mathrm{c.p.}\right]+\frac{6\eta_{2}}{\hbar^{2}m_{0}}\left[\boldsymbol{I}_{1}\boldsymbol{S}_{\mathrm{h}1}\left(\hat{B}_{2}r_{3}p_{1}-\hat{B}_{3}r_{2}p_{1}\right)+\mathrm{c.p.}\right]\nonumber \\
\displaybreak[2]\nonumber \\
& +\frac{3\gamma_{3}}{\hbar^{2}m_{0}}\left[\left\{ \boldsymbol{I}_{1},\,\boldsymbol{I}_{2}\right\} \left(\hat{B}_{2}r_{3}p_{2}-\hat{B}_{1}r_{3}p_{1}+\hat{B}_{3}r_{1}p_{1}-\hat{B}_{3}r_{2}p_{2}\right)+\mathrm{c.p.}\right]\nonumber \\
\displaybreak[2]\nonumber \\
& +\frac{3\eta_{3}}{\hbar^{2}m_{0}}\left[\left(\boldsymbol{I}_{1}\boldsymbol{S}_{\mathrm{h}2}+\boldsymbol{I}_{2}\boldsymbol{S}_{\mathrm{h}1}\right)\left(\hat{B}_{2}r_{3}p_{2}-\hat{B}_{1}r_{3}p_{1}+\hat{B}_{3}r_{1}p_{1}-\hat{B}_{3}r_{2}p_{2}\right)+\mathrm{c.p.}\right],
\\
\displaybreak[1]
\nonumber \\
H_{2} & = \: \frac{1}{8m_{0}}\left(\gamma_{1}'+4\gamma_{2}\right)\left[\hat{\boldsymbol{B}}^{2}\boldsymbol{r}^{2}-\left(\hat{\boldsymbol{B}}\cdot\boldsymbol{r}\right)^{2}\right]+\frac{1}{4\hbar^{2}m_{0}}\left(\eta_{1}+2\eta_{2}\right)\left(\boldsymbol{I}\cdot\boldsymbol{S}_{\mathrm{h}}\right)\left[\hat{\boldsymbol{B}}^{2}\boldsymbol{r}^{2}-\left(\hat{\boldsymbol{B}}\cdot\boldsymbol{r}\right)^{2}\right]\nonumber \\
\displaybreak[2]\nonumber \\
& -\frac{3\gamma_{2}}{4\hbar^{2}m_{0}}\left[\boldsymbol{I}_{1}^{2}\left(\hat{B}_{2}r_{3}-\hat{B}_{3}r_{2}\right)^{2}+\mathrm{c.p.}\right]-\frac{3\eta_{2}}{2\hbar^{2}m_{0}}\left[\boldsymbol{I}_{1}\boldsymbol{S}_{\mathrm{h}1}\left(\hat{B}_{2}r_{3}-\hat{B}_{3}r_{2}\right)^{2}+\mathrm{c.p.}\right]\nonumber \\
\displaybreak[2]\nonumber \\
& -\frac{3\gamma_{3}}{2\hbar^{2}m_{0}}\left[\left\{ \boldsymbol{I}_{1},\,\boldsymbol{I}_{2}\right\} \left(\hat{B}_{2}r_{3}-\hat{B}_{3}r_{2}\right)\left(\hat{B}_{3}r_{1}-\hat{B}_{1}r_{3}\right)+\mathrm{c.p.}\right]\nonumber \\
\displaybreak[2]\nonumber \\
& -\frac{3\eta_{3}}{2\hbar^{2}m_{0}}\left[\left(\boldsymbol{I}_{1}\boldsymbol{S}_{\mathrm{h}2}+\boldsymbol{I}_{2}\boldsymbol{S}_{\mathrm{h}1}\right)\left(\hat{B}_{2}r_{3}-\hat{B}_{3}r_{2}\right)\left(\hat{B}_{3}r_{1}-\hat{B}_{1}r_{3}\right)+\mathrm{c.p.}\right].
\end{align}
\end{widetext}

In our calculations, we express the magnetic field in spherical coordinates [see Eq.~(\ref{eq:spherical_coord})].
For the different orientations of the magnetic field 
we rotate the coordinate system by 
\begin{equation}
\boldsymbol{R}=\left(\begin{array}{ccc}
\cos\varphi\cos\vartheta & \sin\varphi\cos\vartheta & -\sin\vartheta\\
-\sin\varphi & \cos\varphi & 0\\
\cos\varphi\sin\vartheta & \sin\varphi\sin\vartheta & \cos\vartheta
\end{array}\right),
\end{equation}
i.e., we replace $\boldsymbol{x}\rightarrow\boldsymbol{x}'=\boldsymbol{R}^{\mathrm{T}}\boldsymbol{x}$ with $\boldsymbol{x}\in\left\{\boldsymbol{r},\,\boldsymbol{p},\,\boldsymbol{L},\,\boldsymbol{I},\,\boldsymbol{S}\right\}$
to make the quantization axis coincide with the direction of the magnetic
field~\cite{44,ED}.
Finally we express the Hamiltonian in terms of irreducible tensors (see, e.g., Refs.~\cite{ED,7_11,100,125})
and calculate the matrix elements of the matrices $\boldsymbol{D}$ and $\boldsymbol{M}$
in the generalized eigenvalue problem~(\ref{eq:gev}).

\section{Normal vectors\label{sec:normalvec}}

\textcolor{black}{Here we list the normal vectors the nine symmetry planes of the cubic lattice
mentioned in the discussion of Sec.~\ref{sec:analytical}:
\begin{eqnarray}
\hat{\boldsymbol{n}}_{1} & = & \left(1,\,0,\,0\right)^{\mathrm{T}},\nonumber \\
\hat{\boldsymbol{n}}_{2} & = & \left(0,\,1,\,0\right)^{\mathrm{T}},\nonumber \\
\hat{\boldsymbol{n}}_{3} & = & \left(0,\,0,\,1\right)^{\mathrm{T}},\nonumber \\
\hat{\boldsymbol{n}}_{4} & = & \left(1,\,1,\,0\right)^{\mathrm{T}}/\sqrt{2},\nonumber \\
\hat{\boldsymbol{n}}_{5} & = & \left(0,\,1,\,1\right)^{\mathrm{T}}/\sqrt{2},\nonumber \\
\hat{\boldsymbol{n}}_{6} & = & \left(1,\,0,\,1\right)^{\mathrm{T}}/\sqrt{2},\nonumber \\
\hat{\boldsymbol{n}}_{7} & = & \left(1,\,-1,\,0\right)^{\mathrm{T}}/\sqrt{2},\nonumber \\
\hat{\boldsymbol{n}}_{8} & = & \left(0,\,1,\,-1\right)^{\mathrm{T}}/\sqrt{2},\nonumber \\
\hat{\boldsymbol{n}}_{9} & = & \left(-1,\,0,\,1\right)^{\mathrm{T}}/\sqrt{2}.\label{eq:ni}
\end{eqnarray}}



\end{document}